\documentclass[11pt]{article}

\usepackage[T1]{fontenc}
\usepackage[utf8]{inputenc}
\usepackage{lmodern}
\usepackage[margin=1in]{geometry}
\usepackage{microtype}
\usepackage{graphicx}
\usepackage[table]{xcolor}
\usepackage{booktabs}
\usepackage{tabularx}
\usepackage{float}
\usepackage{flafter}
\usepackage{placeins}
\usepackage{amsmath}
\usepackage{amssymb}
\usepackage{tikz}
\usetikzlibrary{arrows.meta,backgrounds,fit,positioning}
\usepackage{pgfplots}
\usepgfplotslibrary{groupplots}
\pgfplotsset{compat=1.18}
\usepackage{url}
\usepackage[numbers,sort&compress]{natbib}
\usepackage{hyperref}
\usepackage{listings}
\usepackage[breakable,skins]{tcolorbox}
\tcbuselibrary{listings}

\definecolor{semaablue}{HTML}{245A7A}
\definecolor{semaalight}{HTML}{EAF2F7}
\definecolor{semaagold}{HTML}{B7791F}
\definecolor{semaagreen}{HTML}{3F7D6B}
\definecolor{modelblue}{HTML}{0072B2}
\definecolor{modelorange}{HTML}{D55E00}
\definecolor{modelpurple}{HTML}{7A5195}
\hypersetup{
  colorlinks=true,
  linkcolor=semaablue,
  citecolor=semaablue,
  urlcolor=semaablue,
  pdfauthor={SEMAADB authors},
  pdftitle={SEMAADB paper draft}
}

\newcommand{\dataset}{\textsc{SEMAADB}}

\newcolumntype{Y}{>{\raggedright\arraybackslash}X}
\lstdefinestyle{paperprompt}{
  basicstyle=\ttfamily\normalsize,
  backgroundcolor=\color{black!3},
  frame=single,
  rulecolor=\color{black!28},
  framesep=6pt,
  breaklines=true,
  breakatwhitespace=true,
  columns=fullflexible,
  keepspaces=true,
  showstringspaces=false,
  aboveskip=7pt,
  belowskip=9pt
}
\newtcolorbox{systemdescription}{
  enhanced,
  breakable,
  colback=black!3,
  colframe=black!28,
  colbacktitle=black!8,
  coltitle=black,
  fonttitle=\bfseries\small,
  fontupper=\small,
  title={System description},
  boxrule=0.4pt,
  arc=0pt,
  outer arc=0pt,
  left=9pt,
  right=9pt,
  top=7pt,
  bottom=7pt,
  before skip=8pt,
  after skip=10pt
}
\newtcblisting{promptbox}[1]{
  enhanced,
  breakable,
  listing only,
  listing engine=listings,
  colback=black!3,
  colframe=black!28,
  colbacktitle=black!8,
  coltitle=black,
  fonttitle=\bfseries\small,
  title={#1},
  title after break={#1\space\textnormal{(continued)}},
  boxrule=0.4pt,
  arc=0pt,
  outer arc=0pt,
  left=6pt,
  right=6pt,
  top=6pt,
  bottom=6pt,
  before skip=8pt,
  after skip=10pt,
  listing options={
    basicstyle=\ttfamily\scriptsize,
    breaklines=true,
    breakatwhitespace=true,
    breakautoindent=false,
    breakindent=0pt,
    columns=fullflexible,
    keepspaces=true,
    showstringspaces=false,
    aboveskip=0pt,
    belowskip=0pt
  }
}
\graphicspath{{figures/}}

\title{A Validated Dataset and Benchmark for Coherent Multi-Diagram SysML Models}

\author{
  \begin{minipage}[t]{0.45\textwidth}
    \centering
    Ardalan Aryashad\\
    University of Southern California\\
    \texttt{aryashad@usc.edu}
  \end{minipage}
  \hfill
  \begin{minipage}[t]{0.45\textwidth}
    \centering
    Yan Jin\\
    University of Southern California\\
    \texttt{yjin@usc.edu}
  \end{minipage}
}

\date{}

\begin{document}

\maketitle

\begin{abstract}
Systems engineers use several diagrams to describe the structure and behavior of systems.
Engineers create these diagrams together to make sure that they use the same elements and remain
consistent with one another. Large language models can generate diagrams as text or code, which
makes it possible to create system diagrams automatically. However, their ability to generate
coherent sets of diagrams is not well understood, and existing datasets and benchmarks do not
directly measure this ability at scale.
We introduce \dataset \: (Systems Engineering Modeling Assistant with AI Dataset and Benchmark) a dataset of 3,000 engineering contexts and 15,000 diagrams. Each context contains five connected SysML views: Requirement, Block Definition, Activity, State Machine, and Sequence.
Here, a view is a diagram that presents one aspect of a system. 
We checked the diagram sets for consistency and valid rendering. A set of 100 contexts is also human-verified and forms the benchmark test set.
We evaluate three language models on two tasks. In diagram repair, the strongest model repairs 64.3\% of semantic errors . In cross-diagram
update the best propagation F1 is 80.7\% when a model applies one change across related diagrams.
The results show that syntax repair is nearly solved, but semantic repair and consistency across diagrams are
still challenging tasks for models.
\dataset{} therefore provides both a large diagram resource and a set of benchmarks for measuring coherent multi-diagram SysML generation.

\end{abstract}

\section{Introduction}
\label{sec:introduction}

Modern engineered systems combine mechanical parts, electronics, software, control logic, and
human interactions. Designing them requires engineers from different fields to share information
about requirements, architecture, behavior, and interfaces throughout the system life cycle.
Model-Based Systems Engineering (MBSE) supports this work by organizing that information in a
connected system model rather than in separate documents alone. The Systems Modeling Language
(SysML) provides a standard way to represent the model through diagrams of requirements,
structure, behavior, and interaction \citep{omg2019sysml}. These diagrams help engineers discuss
a design, trace requirements, find missing information, and study the effect of a change before
the system is built.

In this paper, we define a SysML \emph{view} as a diagram that presents one aspect of a system.
For example, a Requirement view describes what the system must do, a Block Definition view
shows its main components and relations, an Activity view presents a workflow, a State Machine
view shows changes in operating state, and a Sequence view presents interactions over time. Each
view has a different purpose, but the views are not independent. A controller named in a sequence
should match a block in the structural view. A requirement should refer to the correct system
element. States, activities, and messages should also describe the same system behavior. Engineers
therefore create and review the views as one diagram set.

Keeping the views coherent is necessary because an inconsistency can hide a design error or make a
change difficult to trace. Maintaining this coherence is time-consuming and makes verification
challenging because engineers must perform several checks within each view and across the complete
diagram set. They must follow the same names, elements, relations, and requirements across diagrams
that use different notations. When one part of the design changes, they need to update several
views while preserving everything that is not affected. Modeling tools can check some syntax and
formal rules, but many semantic links still depend on the meaning of the system and the intent of
the engineer.

Large language models (LLMs) create a new opportunity for SysML tools. Because a diagram can be
written as text or code, an LLM can turn natural-language requirements into model elements,
generate diagram source, explain a model, or suggest an update. Recent studies have already used
LLMs to generate SysML behavior models from requirements \citep{wang2025generating}, build
computational models through an intermediate SysML representation \citep{hendricks2025text},
align independently developed SysML v2 models \citep{li2025alignment}, and propagate requirement
changes into automotive architectures \citep{cheng2026trustworthy}.
Furthermore the agentic assistants have also been
integrated into MBSE tools \citep{aryashad2026semaa}.
These studies show growing
interest in adding AI assistance to systems-modeling workflows.

This use of LLMs also creates a difficult reliability problem. A generated diagram may look
reasonable while using the wrong element, reversing a relation, omitting a requirement, or
changing behavior. It may also contain valid-looking code that does not render. The problem becomes
harder across several views. The model must understand each notation, preserve exact shared names
and relations, and connect information that appears in different forms. A local edit can require
several related changes, but unnecessary edits can damage parts of the model that were already
correct.

Despite this progress, existing research mainly studies one generated model or one diagram at a
time. SysMBench evaluates requirements-to-model generation on a small human-curated benchmark
\citep{jin2025sysmbench}. Wang et al. provide a small collection of PlantUML behavior models, but
the Activity, State Machine, and Sequence diagrams represent separate systems rather than
connected views of the same system \citep{wang2025generating}. Larger synthetic UML collections
also focus on individual diagram generation \citep{bates2025uml}. As a result, current resources
offer limited support for training or studying tools that must work with several related SysML
diagrams.

To our knowledge, there is no established benchmark for judging whether an AI tool preserves the
meaning and consistency of a complete diagram set. Rendering alone is not enough: code may compile
even when a relation is reversed or a required element is missing. Testing one diagram alone also
cannot show whether a change was carried into every related view. The field therefore needs both
a large collection of connected diagrams and evaluation tasks that measure semantic repair and
cross-diagram consistency. These resources are necessary for developing AI-assisted modeling tools,
comparing them fairly, and identifying the errors that still require engineering review.

These limitations motivated \dataset which the main contributions of this work are:
\begin{enumerate}
  \item \textbf{A large multi-diagram SysML dataset.} We provide 3,000 engineering contexts and
  15,000 diagrams covering Requirement, Block Definition, Activity, State Machine, and Sequence
  views across four engineering domains.
  \item \textbf{A pipeline for coherent diagram-set generation.} We use a shared entity
  description to connect the five views, then check consistency and rendering while preserving
  generation provenance.
  \item \textbf{A human-verified benchmark set.} We human-verify 100 contexts under a documented
  protocol and use only this set for the evaluation.
  \item \textbf{Benchmarks for repair and cross-diagram updates.} The tasks measure whether a
  model can correct semantic errors and apply a change across related diagrams without damaging
  unaffected content.
  \item \textbf{An objective evaluation of three language models.} We report task scores, render
  validity, confidence intervals, failure patterns, and inference cost, showing where current
  models succeed and where coherent SysML modeling remains challenging.
\end{enumerate}

The dataset is available at \href{https://huggingface.co/datasets/ardalon/semaadb-dataset}{huggingface.co/datasets/ardalon/semaadb-dataset}.

\section{Related Work}
\label{sec:related-work}

\subsection{AI and agentic assistance for SysML}

AI-assisted modeling has moved from extracting model elements to creating and updating formal
models. Wang et al. generate SysML behavior diagrams from natural-language requirements
\citep{wang2025generating}. Hendricks and Cicirello use SysML as an intermediate step between
engineering documents and computational models \citep{hendricks2025text}. SEMAA uses specialized
agents to generate and check diagrams within an MBSE tool \citep{aryashad2026semaa}. Other work
applies LLMs to semantic alignment between SysML v2 models \citep{li2025alignment} and to
requirement-driven architecture updates in automotive systems \citep{cheng2026trustworthy}.
Together, these studies show that LLMs can support several parts of a modeling workflow, including
agentic workflows with explicit intermediate representations. However, they focus on a specific
model, diagram, or case study. Data and benchmarks are still needed to test whether artifacts
remain correct as a modeling workflow becomes longer and more connected.

\subsection{Automatic data generation}

Automatic data generation is widely used when manual creation is expensive. In code generation,
Magicoder creates instruction data from open-source code and uses those examples to train smaller
code models \citep{wei2024magicoder}. In vision-language research, CoSyn asks language models to
write programs that render text-rich images and then derives instruction data from the same code
\citep{yang2025cosyn}. These methods show that generated data can reach useful scale when its
structure is controlled and its outputs can be checked.
Diagram and chart research follows a similar pattern. ChartCoder builds chart-to-code data through
generation and execution filtering \citep{zhao2025chartcoder}. StarFlow combines synthetic,
annotated, and real workflow diagrams for structured workflow generation
\citep{bechard2025starflow}. OmniDiagram uses a large generated collection spanning several
diagram languages and tasks \citep{yang2026omnidiagram}. Across these projects, code is valuable
because it can be executed or rendered, providing a low-cost quality gate before human review.

This pattern supports automatic generation for a SysML dataset, but it does not remove the need
for domain checks. A rendered chart can be compared with an image, while a SysML diagram must also
match engineering requirements and agree with other diagrams of the same system. Our pipeline
therefore adds a shared system representation and cross-diagram checks to the common
generation, rendering, and filtering process. Human review is then concentrated on the benchmark
test set, where semantic quality has the greatest effect on reported results.

\begin{table}[t]
  \centering
  \caption{Comparison with the closest datasets and benchmarks}
  \label{tab:related-work-comparison}
  \small
  \begin{tabularx}{\linewidth}{@{}l Y c c c@{}}
    \toprule
    \textbf{Dataset} & \textbf{Main content} & \textbf{SysML} &
    \textbf{Linked views} & \textbf{Evaluation} \\
    \midrule
    SysMBench \citep{jin2025sysmbench} & Requirements to textual system models & $\checkmark$ & $\times$ & $\times$ \\
    Wang et al. \citep{wang2025generating} & PlantUML behavior diagrams & $\checkmark$ & $\times$ & $\times$ \\
    Bates et al. \citep{bates2025uml} & Synthetic UML image-to-code pairs & $\times$ & $\times$ & $\times$ \\
    OmniDiagram \citep{yang2026omnidiagram} & General diagram-code tasks & $\times$ & $\times$ & $\times$ \\
    \textbf{\dataset{}} & \textbf{Five connected views for each system} &
    $\boldsymbol{\checkmark}$ & $\boldsymbol{\checkmark}$ & $\boldsymbol{\checkmark}$ \\
    \bottomrule
  \end{tabularx}
\end{table}

\subsection{Benchmarks and evaluation}

Existing SysML resources establish important baselines but cover a narrower unit of data.
SysMBench evaluates the generation of SysML v2 models from requirements on a small human-curated
benchmark \citep{jin2025sysmbench}.
Bates et al. create a much larger synthetic UML collection for image-to-code generation, focused
on Activity and Sequence diagrams \citep{bates2025uml}. These datasets test whether a model can
produce one model or recover one diagram, not whether several SysML views remain coherent.
General diagram benchmarks cover more tasks but retain the same single-artifact focus. OmniDiagram
includes diagram-to-code, text-to-code, and diagram-editing tasks \citep{yang2026omnidiagram}.
FlowVQA derives questions from flowchart structure to test visual and logical reasoning
\citep{singh2024flowvqa}. These resources evaluate diagram
syntax, perception, structure, or local reasoning, but they do not test several connected SysML
views of the same system or the propagation of a change across those views.
Generated data is also used directly for model evaluation in other fields. In code generation,
EvoEval uses an LLM to transform existing programming problems into targeted benchmark variants
\citep{xia2024evoeval}. In vision, researchers also use rendered scenes application-relevant measures to evaluate model performance 
and  diagnose visual reasoning abilities \citep{johnson2017clevr, aryashad2025filters}.
Together,
these benchmarks show that generated data can provide scale and controlled coverage when paired
with clear generation rules, quality checks, and measures that reflect the intended task.

SEMAADB follows both ideas for systems modeling. It uses generated contexts to cover varied systems
and reserves a human-verified subset for benchmark evaluation. For SysML, a rendered output is
insufficient because the engineering content and relations must also be correct. We therefore
separate render validity from semantic scoring and use exact repair and graph-derived update
measures for the two headline tasks. This combination of linked views and set-level evaluation is
the gap addressed by \dataset. Table \ref{tab:related-work-comparison} summarize the comparison of our work with other related datasets

\section{Dataset Construction}
\label{sec:dataset}

\begin{figure*}[t]
  \centering
  \resizebox{0.98\textwidth}{!}{%
    \begin{tikzpicture}[
      font=\small,
      node distance=5mm,
      stage/.style={
        draw=semaablue,
        fill=white,
        rounded corners=2pt,
        align=center,
        minimum height=13mm,
        minimum width=34.5mm,
        text width=29mm,
        inner sep=4pt
      },
      validation/.style={
        stage,
        minimum width=98mm,
        text width=91mm
      },
      output/.style={
        stage,
        minimum width=50mm,
        text width=43mm
      },
      review/.style={
        stage,
        minimum width=47.67mm,
        text width=40mm
      },
      panel/.style={
        rounded corners=3pt,
        minimum width=159mm,
        minimum height=23mm
      },
      generation panel/.style={
        panel,
        draw=semaablue!35,
        fill=semaablue!6
      },
      automatic panel/.style={
        panel,
        draw=semaagold!40,
        fill=semaagold!7
      },
      human panel/.style={
        panel,
        draw=semaagreen!40,
        fill=semaagreen!7
      },
      phase/.style={
        font=\bfseries,
        align=right,
        text width=25mm
      },
      flow/.style={-{Latex[length=2.2mm]}, draw=semaablue, line width=0.9pt}
    ]
      \node[stage, anchor=west] (coverage) at (0,0) {\textbf{1. Coverage}\\Domains, systems, and sources};
      \node[stage, right=of coverage] (description) {\textbf{2. Description}\\Grounded system description};
      \node[stage, right=of description] (anchor) {\textbf{3. Shared model}\\Names, types, and relations};
      \node[stage, right=of anchor] (views) {\textbf{4. Diagrams set}\\Connected SysML views};

      \node[output, anchor=west] (automatic) at (0,-28mm) {\textbf{6. Save passed views}\\Code, Render, metadata};
      \node[validation, right=of automatic] (checks) {\textbf{5. Automatic validation}\\Check the context, render and repair each diagram, and compare shared content across the set};

      \node[review, anchor=west] (selection) at (0,-56mm) {\textbf{7. Select eval sets}\\Different domain, type, difficulty, and source};
      \node[review, right=of selection] (humanreview) {\textbf{8. Human verification}\\Review, score, correct};
      \node[review, right=of humanreview] (verified) {\textbf{9. Verified eval set}\\Accepted examples};

      \node[phase, text=semaablue, anchor=east] at (-6mm,0) {Generation};
      \node[phase, text=semaagold, anchor=east] at (-6mm,-28mm) {Automatic\\validation};
      \node[phase, text=semaagreen, anchor=east] at (-6mm,-56mm) {Human\\verification};

      \begin{scope}[on background layer]
        \node[generation panel] at (76.5mm,0) {};
        \node[automatic panel] at (76.5mm,-28mm) {};
        \node[human panel] at (76.5mm,-56mm) {};
      \end{scope}

      \coordinate (turntop) at ([xshift=5mm]views.east);
      \coordinate (turnbottom) at ([xshift=5mm]checks.east);

      \draw[flow] (coverage) to (description);
      \draw[flow] (description) to (anchor);
      \draw[flow] (anchor) to (views);
      \draw[flow, rounded corners=3pt]
        (views.east) to (turntop) to (turnbottom) to (checks.east);
      \draw[flow] (checks) to (automatic);
      \draw[flow] (automatic) to (selection);
      \draw[flow] (selection) to (humanreview);
      \draw[flow] (humanreview) to (verified);
    \end{tikzpicture}%
  }
  \caption{Dataset pipeline for generation, automatic validation, and human verification.}
  \label{fig:dataset-pipeline}
\end{figure*}
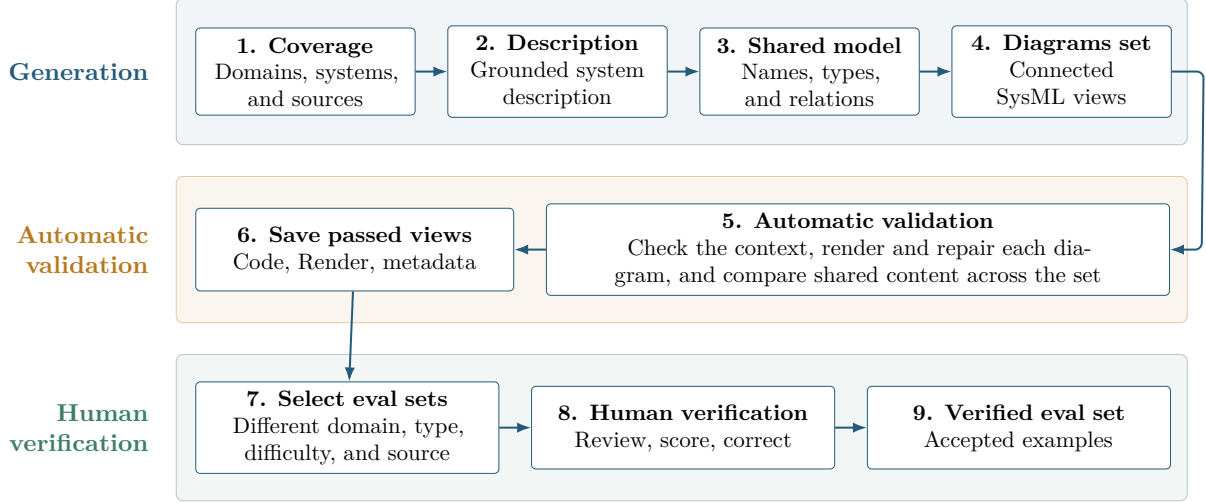

The dataset was created in three phases: defining systems, generating each
diagram set, and validating the generated views. A selected subset was then reviewed by human
annotators. Figure~\ref{fig:dataset-pipeline} summarizes this process.

\subsection{Dataset unit and coverage}

We define a \emph{context} as one engineering system or subsystem. Each context contains a natural
language system description, source metadata, a shared entity model, five diagram sources, their
rendered images, and generation provenance. The five diagrams are the Requirement, Block
Definition, Activity, State Machine, and Sequence views. Together, they describe the needs,
structure, processes, states, and interactions of the same system.

Although each diagram serves a different purpose, all five must describe the same system. We first
create a shared list of the system's parts, requirements, states, activities, actors, messages, and
relationships. When information appears in more than one diagram, its name and meaning come from
this list. This gives each diagram the information it needs while keeping repeated content
consistent across the set.

Before generation, we prepared a plan to cover a broad range of engineering systems. The plan
includes mechanical, electrical, computer  embedded, and mechatronic  control
systems. Within these domains, we selected familiar examples such as transmissions,
battery-management systems, and flight controllers. We also represented systems at the component,
subsystem, and complete-system levels, with changes in their intended use and design goals. This
planned distribution provides varied examples and reduces unnecessary repetition.

\begin{figure}[t]
  \centering
  \begin{tikzpicture}[x=1cm,y=1cm,font=\small]
    \def\domainbar#1#2#3#4{%
      \filldraw[draw=#4,fill=#4!72] (#1-0.34,0.55) rectangle (#1+0.34,{0.55+#3*0.003});
      \node[anchor=south,font=\scriptsize] at (#1,{0.66+#3*0.003}) {#3};
      \node[anchor=north,align=center,font=\scriptsize] at (#1,0.42) {#2};
    }
    \def\entitybar#1#2#3#4{%
      \node[anchor=east,font=\scriptsize] at (10.15,#1) {#2};
      \filldraw[draw=#4,fill=#4!72] (10.32,#1-0.14) rectangle ({10.32+#3*0.47},#1+0.14);
      \node[anchor=west,font=\scriptsize] at ({10.43+#3*0.47},#1) {#3};
    }

    \node[font=\bfseries] at (4.00,4.02) {(a) Contexts by domain};
    \foreach \y/\label in {0.55/0,1.45/300,2.35/600,3.25/900} {
      \draw[black!13] (0.60,\y) to (7.38,\y);
      \node[anchor=east,font=\scriptsize] at (0.48,\y) {\label};
    }
    \draw (0.60,0.55) to (0.60,3.25);
    \domainbar{1.25}{Mechanical}{846}{semaablue}
    \domainbar{3.15}{Electrical}{846}{semaagold}
    \domainbar{5.05}{Computer\\embedded}{843}{semaagreen}
    \domainbar{6.95}{Mechatronic\\control}{465}{modelpurple}

    \node[font=\bfseries] at (12.67,4.02) {(b) Median entities per context};
    \foreach \n in {0,2,4,6,8,10} {
      \draw[black!13] ({10.32+\n*0.47},0.55) to ({10.32+\n*0.47},3.55);
      \node[anchor=north,font=\scriptsize] at ({10.32+\n*0.47},0.42) {\n};
    }
    \draw (10.32,0.55) to (15.02,0.55);
    \entitybar{3.35}{Blocks}{10}{modelblue}
    \entitybar{2.83}{Messages}{10}{modelpurple}
    \entitybar{2.31}{Activities}{8}{semaagreen}
    \entitybar{1.79}{States}{7}{cyan!55!black}
    \entitybar{1.27}{Requirements}{6}{semaagold}
    \entitybar{0.75}{Actors}{2}{modelorange}
  \end{tikzpicture}
  \caption{Coverage and shared-model content of the 3,000 contexts. The left chart
  reports domain coverage. The right chart reports median entity counts.}
  \label{fig:dataset-statistics}
\end{figure}
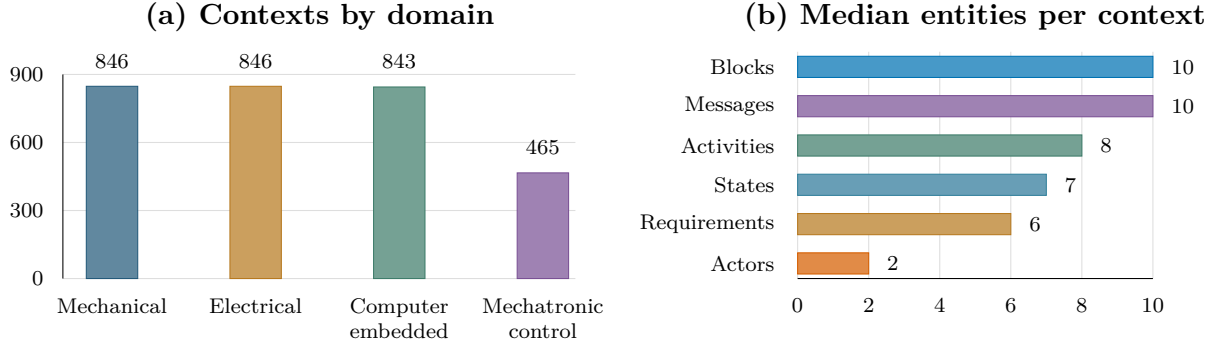

Figure~\ref{fig:dataset-statistics} summarizes the dataset. Mechanical, electrical, and
computer and embedded systems each contribute slightly more than one quarter of the contexts, with
the remaining contexts covering mechatronic and control systems. Component, subsystem, and system
scales are nearly balanced, with about one third of the dataset at each scale. A typical shared
model contains about ten blocks and messages, together with requirements, states, activities, and
actors that connect the five diagrams. The figure uses median counts so that unusually large or
small contexts do not dominate this summary. Detailed statistics are reported in
Appendix~\ref{app:dataset-statistics}.

We searched the internet for public engineering references and terminology related to each planned
system. These sources guided the descriptions, but their content was paraphrased rather than copied.
Each context keeps its source references and license information when available.

\subsection{Generation pipeline}

For each planned system, the pipeline begins with its domain, archetype, scale, and a short list of
seed terms. An LLM with access to the internet creates the system description based on the source
information. The description follows a fixed format so it can be checked before diagram generation.
We used GPT-5.5
to generate the descriptions, shared entity models, and final diagrams. The details of generation were saved for every context.

A second step turns the description into the shared list used by all five diagrams. Before accepting
the list, we check that every item has a distinct name, every relationship connects known items, and
there is enough information to create each diagram. If a check fails, the model receives a clear
explanation and produces a corrected version.

The five diagrams are then created one at a time. For each diagram, the model receives the
system description, the relevant parts of the shared list, and simple rules for that view. It must use the given names without changing them. Creating the diagrams
separately keeps each prompt focused, while the shared list keeps the complete set consistent.
Detailed prompt samples and the instructions for each view are provided in
Appendix~\ref{app:prompts}.

Each context includes the system description, shared list, diagram code, renderer version,
prompt versions, model names, generation time, token use, repair attempts, and validation
results. The PlantUML code and its rendering are stored with this information. This allows an
example to be inspected, rendered again, and traced to its generation settings.

\subsection{Automatic validation}

Automatic validation is applied before a context is saved. The process
uses five checks:

\begin{enumerate}
  \item \textbf{Check that the context is complete.} The system description, source information,
  shared elements, and all five diagrams must be present. Missing content is completed before the
  context can continue.
  \item \textbf{Check shared names and connections.} Names must be unique, and every connection
  must refer to an element that exists. If this check fails, the shared model is corrected and
  checked again.
  \item \textbf{Render every diagram.} Each source code must produce a valid rendered image with
  the PlantUML engine \citep{plantuml}. If rendering fails, the error and diagram source are returned
  to the model for correction, with up to three attempts.
  \item \textbf{Check the content of each diagram.} Each view must contain the expected kinds of
  elements. For example, requirements belong in the Requirement view and states belong in the State
  Machine view.
  \item \textbf{Compare the five diagrams.} Names and relationships that appear in several views
  must agree. For example, a Sequence participant should also appear as a system block or a defined
  actor.
\end{enumerate}

\begin{table}[t]
  \centering
  \caption{Rubric used to verify each selected five-diagram set.}
  \label{tab:human-verification-rubric}
  \small
  \begin{tabularx}{\linewidth}{@{}l Y l@{}}
    \toprule
    \textbf{Criterion} & \textbf{Question} & \textbf{Scale} \\
    \midrule
    Rendering & Do all five diagrams compile and display correctly? & Pass or fail \\
    SysML correctness & Does each view use suitable elements and relations? & 0-3 \\
    Domain plausibility & Are the components and interactions reasonable for the system? & 0-3 \\
    Cross-diagram consistency & Do shared names and relations agree across the five views? & 0-3 \\
    Completeness & Does the set represent the system description? & 0-3 \\
    \bottomrule
  \end{tabularx}
\end{table}

The \texttt{auto-passed} label means that the context is complete, all five diagrams render, and
the content and consistency checks have been completed. It does not mean that a person has reviewed
the engineering content.

\subsection{Human verification}

Human review is performed on complete five-diagram sets rather than isolated diagrams. We selected
100 contexts for the benchmark core using deterministic stratification across domain, archetype,
difficulty, and generation provenance. This selection supports balanced evaluation while keeping
the larger automatically checked portion of the dataset separate.
Before formal annotation, both reviewers completed the same five practice sets and discussed the
rubric. During annotation, a reviewer reads the system description, inspects the shared entity
model, renders all five diagrams, and views the full set together. The reviewer scores diagram based on the four criteria defined in Table \ref{tab:human-verification-rubric}.
The criteria include SysML correctness, domain plausibility,
cross-diagram consistency, and completeness. A score of 0 indicates failure, 1 a major issue, 2 a
minor issue, and 3 a clean output.

Twenty of the selected contexts were assigned independently to both reviewers. Their reviews
remained hidden from each other until both reviews were submitted. The other contexts received one
review, balanced between the reviewers. Disagreements in the shared subset were preserved and then
adjudicated. This design allows agreement to be measured without letting one review influence the
other.
Reviewers could accept a set, make a minimal correction and accept it, or reject it. Corrections
were limited to local problems such as terminology, relation direction, or inconsistent names. A
set that required a redesign was replaced instead of being rewritten. Every corrected artifact was
rendered and checked again, and each change was stored with the original context and reviewer notes.
Reviewed and accepted sets received the \texttt{human-verified} label.

\section{Benchmark Design}
\label{sec:benchmark}

The benchmark measures two abilities that are important when an AI assistant works with a set of
related system diagrams. The diagram repair benchmark measures whether a model can correct one
damaged diagram while preserving its valid content. The cross-diagram update benchmark measures
whether a model can apply a change consistently across other related diagram. The following
subsections describe the inputs, expected outputs, and evaluation measures for both benchmarks.

Benchmarks use the same 100 contexts accepted during human verification, as described in
Section~\ref{sec:dataset}. We fixed this evaluation set before testing the models and used it
unchanged in every experiment. Therefore, all models are evaluated on the same systems and the
same diagram set. Each result is linked to its verified context and benchmark example, which
makes the evaluation traceable and repeatable.
We also explored diagram-to-text description and diagram-to-diagram translation. These benchmarks
need further work on their task design and evaluation, so their current details are provided in
Appendix~\ref{app:development-benchmarks}.

\subsection{Diagram repair benchmark}
\label{sec:task-repair}

The diagram repair benchmark presents the model with a system description and all five diagrams from one
context. Exactly one diagram has been changed and the other four remain unchanged. The model must
identify the problem and return the complete repaired views for the damaged diagram. The prompt
asks the model to change only the damaged content and preserve all other elements and relations.

We create two categories of damage. The \emph{syntax} category contains changes that prevent the
diagram from rendering, such as a damaged arrow, keyword, stereotype, brace, declaration, or
quotation mark. The renderer message is included in the prompt. The \emph{semantic} category remains
renderable but conflicts with the system description or the other diagrams. Its changes rename an
entity, delete an edge, reverse or redirect an edge, remove a requirement, or exchange two entity
names. No renderer message is provided for this category because the diagram is syntactically valid.

Each operator is applied once per context when a suitable element is available. We retain a syntax
instance only if the changed diagram fails to render and a semantic instance only if it still
renders and changes the extracted model. This process produces 557 syntax instances and 600
semantic instances, for 1,157 repair instances in total.

The primary measure is \emph{exact repair}. We convert the original, damaged, and returned
diagrams into typed graphs of elements and relations. The damaged part must match the original
graph after repair, and every part outside that region must remain unchanged. A repair is counted
as exact only when it satisfies both conditions. We call the first condition adherence and the
second preservation. Before comparison, the arrow directions and aliases are
normalized so that formatting and declaration order do not affect the score. A response that
cannot be parsed or rendered receives an exact-repair score of zero.

\subsection{Cross-diagram update benchmark}
\label{sec:task-update}

The cross-diagram update benchmark tests whether a model can carry one change from an edited diagram to the other
diagrams that describe the same system. The model receives the system description, all five
diagrams, and a description of the edit. It must
select the affected diagrams, explain why they require changes, and return valid diagram. We use three categories of edits. \emph{Rename} changes the name of an element shared by several diagrams.
\emph{Remove} deletes an element and requires the removal of relations that would otherwise become
stale. \emph{Add} introduces a connected element and allows more than one valid placement in
the related diagrams. Each context contributes one instance from every category, which produces
300 update instances.

Expected updates are derived from the shared model and the typed graphs of the five diagrams. Each
required change is represented by a tuple containing the diagram, element, and operation. A true
positive is a required change that appears in the returned graph. A false negative is a missed
change, and a false positive is an unrelated structural change. We compute precision, recall, and
the F1 score as follows:
\begin{equation}
  P = \frac{\mathrm{TP}}{\mathrm{TP}+\mathrm{FP}}, \qquad
  R = \frac{\mathrm{TP}}{\mathrm{TP}+\mathrm{FN}}, \qquad
  F_1 = \frac{2PR}{P+R},
  \label{eq:propagation-f1}
\end{equation}
where $P$ is precision, $R$ is recall, $\mathrm{TP}$ is the number of correct updates,
$\mathrm{FP}$ is the number of unrelated changes, and $\mathrm{FN}$ is the number of missed
updates.
$F_1$ score is the primary measure.

\begin{table}[H]
  \centering
  \caption{Main benchmark results. Uncertainty is shown as the upper and lower 95\% confidence
  range in percentage points. The best result in each row is bold.}
  \label{tab:headline-results}
  \small
  \begin{tabularx}{\textwidth}{@{}Ycccc@{}}
    \toprule
    \textbf{Benchmark measure} & \textbf{Examples} & {\textbf{Luna}} & {\textbf{Terra}} &
    {\textbf{Sol}} \\
    \midrule
    Semantic exact repair & 600 & {$48.3^{+3.4}_{-3.3}\%$} & {$58.5^{+3.0}_{-3.0}\%$} &
    {\boldmath$\mathbf{64.3}^{+3.0}_{-3.0}\%$} \\
    Syntax exact repair & 557 & {$98.0^{+1.1}_{-1.1}\%$} & {$99.1^{+0.7}_{-0.9}\%$} &
    {\boldmath$\mathbf{99.8}^{+0.2}_{-0.3}\%$} \\
    Cross-diagram update $F_1$ & 300 & {$73.0^{+2.5}_{-2.4}\%$} & {$77.5^{+2.7}_{-2.8}\%$} &
    {\boldmath$\mathbf{80.7}^{+2.7}_{-2.8}\%$} \\
    \bottomrule
  \end{tabularx}
\end{table}

\begin{figure}[H]
  \centering
  \begin{tikzpicture}
    \begin{groupplot}[
      group style={group size=2 by 1, horizontal sep=16mm},
      width=0.47\textwidth,
      height=61mm,
      ymajorgrids,
      grid style={draw=black!10},
      axis line style={draw=black!55},
      tick style={draw=black!55},
      tick label style={font=\small},
      label style={font=\small},
      title style={font=\bfseries},
      legend style={
        at={(1.12,1.28)},
        anchor=south,
        legend columns=3,
        draw=none,
        font=\small
      }
    ]
      \nextgroupplot[
        title={Diagram repair},
        ylabel={Exact repair (\%)},
        xmin=0.55,
        xmax=2.45,
        ymin=40,
        ymax=105,
        xtick={1,2},
        xticklabels={Syntax,Semantic},
        ytick={40,50,60,70,80,90,100}
      ]
        \addplot+[
          only marks, mark=*, mark size=2.6pt, color=modelblue,
          error bars/.cd, y dir=both, y explicit,
          error bar style={line width=0.8pt},
          error mark options={rotate=90, mark size=3pt, line width=0.8pt}
        ] coordinates {
          (0.85,98.0) +- (0,1.1)
          (1.85,48.3) +- (0,3.4)
        };
        \addlegendentry{GPT-5.6 Luna}
        \addplot+[
          only marks, mark=square*, mark size=2.6pt, color=modelorange,
          error bars/.cd, y dir=both, y explicit,
          error bar style={line width=0.8pt},
          error mark options={rotate=90, mark size=3pt, line width=0.8pt}
        ] coordinates {
          (1.00,99.1) +- (0,0.8)
          (2.00,58.5) +- (0,3.0)
        };
        \addlegendentry{GPT-5.6 Terra}
        \addplot+[
          only marks, mark=triangle*, mark size=2.8pt, color=modelpurple,
          error bars/.cd, y dir=both, y explicit,
          error bar style={line width=0.8pt},
          error mark options={rotate=90, mark size=3pt, line width=0.8pt}
        ] coordinates {
          (1.15,99.8) +- (0,0.3)
          (2.15,64.3) +- (0,3.0)
        };
        \addlegendentry{GPT-5.6 Sol}

      \nextgroupplot[
        title={Cross-diagram update},
        ylabel={Propagation $F_1$ (\%)},
        xmin=0.55,
        xmax=4.45,
        ymin=25,
        ymax=105,
        xtick={1,2,3,4},
        xticklabels={Rename,Removal,Addition,Overall},
        ytick={30,40,50,60,70,80,90,100}
      ]
        \addplot+[
          only marks, mark=*, mark size=2.6pt, color=modelblue,
          error bars/.cd, y dir=both, y explicit,
          error bar style={line width=0.8pt},
          error mark options={rotate=90, mark size=3pt, line width=0.8pt}
        ] coordinates {
          (0.85,97.8) +- (0,2.5)
          (1.85,81.7) +- (0,2.4)
          (2.85,39.6) +- (0,7.1)
          (3.85,73.0) +- (0,2.5)
        };
        \addplot+[
          only marks, mark=square*, mark size=2.6pt, color=modelorange,
          error bars/.cd, y dir=both, y explicit,
          error bar style={line width=0.8pt},
          error mark options={rotate=90, mark size=3pt, line width=0.8pt}
        ] coordinates {
          (1.00,100.0) +- (0,0.0)
          (2.00,92.2) +- (0,2.3)
          (3.00,40.3) +- (0,8.4)
          (4.00,77.5) +- (0,2.8)
        };
        \addplot+[
          only marks, mark=triangle*, mark size=2.8pt, color=modelpurple,
          error bars/.cd, y dir=both, y explicit,
          error bar style={line width=0.8pt},
          error mark options={rotate=90, mark size=3pt, line width=0.8pt}
        ] coordinates {
          (1.15,100.0) +- (0,0.0)
          (2.15,99.3) +- (0,0.7)
          (3.15,42.7) +- (0,8.4)
          (4.15,80.7) +- (0,2.8)
        };
    \end{groupplot}
  \end{tikzpicture}
  \caption{Model performance on diagram repair and cross-diagram update. Points show the mean
  percentage and line show 95\% confidence intervals.}
  \label{fig:benchmark-results}
\end{figure}
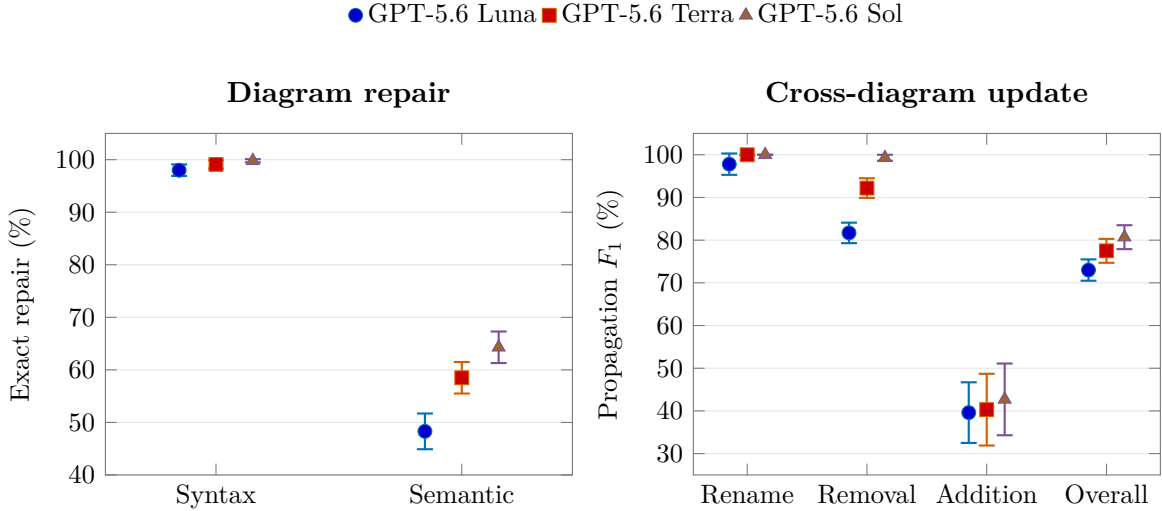

\subsection{Models and inference}

We evaluate GPT-5.6 Luna, GPT-5.6 Terra, and GPT-5.6 Sol. All models received the same benchmark
examples, prompts, output format, and inference settings; only the model was changed. We collected
one response for each example. Both prompts were zero-shot and required one structured response
containing complete diagrams rather than patches or partial code.
The evaluation contains 4,371 responses: 1,157 repair and 300 update responses from each of the
three models. Before scoring model outputs, a no-change response and a reference response were used
as sanity checks for each scoring method.

\subsection{Statistical analysis}

For diagram repair, model performance is the percentage of cases repaired exactly. The semantic and syntax repair scores are reported
separately. For cross-diagram update, we calculate $F_1$ for each example and
then average these scores across the 300 examples.
We also report 95\% confidence intervals to show how much the scores vary across the systems in the test
set. These intervals are calculated by resampling complete contexts and recalculating the model
scores. Keeping all examples from the same context together accounts for their shared system
content.

\FloatBarrier
\section{Results}
\label{sec:results}

The benchmarks reveal two clear patterns. The models repair syntax reliably, but semantic repair
remains difficult. They also handle renaming and removal more successfully than adding new
information across diagrams. Table~\ref{tab:headline-results} presents the main results and their
95\% confidence intervals. Figure~\ref{fig:benchmark-results} compares performance across the
benchmark categories. All performance values are reported as percentages, and higher is better.
Running the complete headline evaluation cost \$37.40, as reported by the OpenAI platform.

\subsection{Diagram repair}

All three models nearly solve the syntax category. Exact-repair rates are 98.0\% for Luna, 99.1\%
for Terra, and 99.8\% for Sol. Because these results are close to saturation, we use the semantic
category as the main repair result.

Semantic repair is substantially harder. Luna repairs 48.3\% of the examples exactly, Terra
repairs 58.5\%, and Sol repairs 64.3\%. Even the strongest model fails to produce an
exact repair of the semantic examples. When both categories are combined, exact-repair
rates are 72.3\%, 78.0\%, and 81.4\% for Luna, Terra, and Sol, respectively. The combined values are
not treated as the main result because the easier syntax examples would hide part of the semantic
difficulty.

Table~\ref{tab:repair-breakdown} shows that performance depends strongly on the type of damage.
Redirecting and reversing relations are the easiest changes for Terra and Sol. Renaming an entity
is also handled relatively well. Removing a requirement is the hardest category, with exact-repair
rates between 8.0\% and 9.0\%. Deleting an edge and exchanging two names are also difficult. This
pattern suggests that models handle direct substitutions more reliably than changes requiring
careful removal or the preservation of two identities. Paired comparisons support the performance
ordering of Sol, Terra, and Luna on semantic repair.

\begin{table}[!h]
  \centering
  \caption{Exact-repair rate for each semantic damage category.}
  \label{tab:repair-breakdown}
  \small
  \begin{tabular}{@{}lccc@{}}
    \toprule
    \textbf{Damage category} & \textbf{Luna} & \textbf{Terra} & \textbf{Sol} \\
    \midrule
    Delete edge & 37.0\% & 32.0\% & \textbf{40.0\%} \\
    Remove requirement & 8.0\% & 8.0\% & \textbf{9.0\%} \\
    Rename entity & 70.0\% & 81.0\% & \textbf{84.0\%} \\
    Redirect edge & 76.0\% & \textbf{94.0\%} & \textbf{94.0\%} \\
    Reverse edge & 77.0\% & 87.0\% & \textbf{92.0\%} \\
    Exchange names & 22.0\% & 49.0\% & \textbf{67.0\%} \\
    \bottomrule
  \end{tabular}
\end{table}

\subsection{Cross-diagram update}

The overall $F_1$ score is 73.0\% for Luna, 77.5\% for Terra, and 80.7\% for Sol, and all 900 model
responses in this benchmark render successfully

The edit categories reveal a large difference in difficulty. For renaming, Luna reaches 97.8\% and
both Terra and Sol reach 100.0\%. For removal, performance remains high but the models separate more
clearly: 81.7\% for Luna, 92.2\% for Terra, and 99.3\% for Sol. Addition is much harder, with scores
of 39.6\%, 40.3\%, and 42.7\%. Thus, the models can usually propagate a change to existing content, but
they often fail to place and connect new information across the relevant diagrams.

The additional measures help explain this result. Diagram selection is similar for all three
models: 84.7\% for Luna, 83.5\% for Terra, and 84.8\% for Sol. The difference appears in what the
models change after selecting the diagrams. Unrelated edits occur in 9.5\%, 3.3\%, and 0.3\% of
the evaluated changes, respectively. In addition, 20.0\% of Luna responses, 5.7\% of Terra
responses, and 1.0\% of Sol responses leave a mismatch between diagrams. Sol therefore gains its
advantage mainly by avoiding unnecessary changes and resolving more cross-diagram inconsistencies.
Paired comparisons support the performance ordering of Sol, Terra, and Luna on the update
benchmark.

\section{Discussion and Ablation Studies}
\label{sec:analysis}

\subsection{Interpretation of Benchmark Performance}

The large difference between syntax and semantic repair shows that generating valid code is not the main
difficulty. A model can often use a renderer message to correct a malformed arrow or delimiter.
Semantic repair instead requires the model to compare several sources of evidence, identify what
is missing or incorrect, and avoid changing valid content. This is especially difficult when the
error is an absence. A missing requirement or relation leaves no incorrect statement for the model
to edit directly. This helps explain why removing a requirement from the input was much harder to
repair than renaming an entity or changing the direction of a visible relation.

The update results show a similar pattern. Renaming and removal begin with an element that already
appears in the diagram set. Addition requires the model to infer where new information should
appear and how it should connect to existing content. Because diagram-selection scores are similar,
the performance differences arise mainly after a diagram has been selected. Stronger performance
comes from applying the intended change precisely, avoiding unrelated edits, and resolving the
dependencies between diagrams.

These findings also show why both benchmarks are needed. Diagram repair measures whether a model
can recover one intended representation while preserving its surroundings. Cross-diagram update
allows several valid representations, but requires a change to remain coherent across the set.
Together, they separate rendering ability from system-level reasoning.

\subsection{Failure Analysis}

Two examples illustrate the main failure patterns. In one repair example, the requirement for an
ABS controller's supply-voltage range was removed from the Requirement diagram. A Sol response did
not restore the missing requirement. Instead, it reversed an existing relation between two other
requirements. The result rendered correctly, but it repaired a visible relation that was already
valid and left the actual problem unresolved. This example shows that a plausible local edit can
still be inconsistent with the system description and the other diagrams.

In one update example, an auxiliary solenoid-inlet-valve block was added to the Block Definition
diagram. The expected update also introduced the element in the Requirement and Sequence diagrams.
All three models returned no additional diagrams. Their responses treated the new block as local to
the Block Definition diagram because its name did not yet appear elsewhere. The models therefore
used current references as evidence, but did not infer the new requirement and interaction implied
by the system change. More information about these cases is provided in
Appendix~\ref{app:failure-examples}.

\begin{figure}[H]
  \centering
  \setlength{\fboxsep}{0pt}

  \fcolorbox{black!28}{white}{%
    \begin{minipage}{0.965\textwidth}
      \colorbox{black!8}{%
        \parbox{\dimexpr\linewidth-2\fboxsep\relax}{%
          \vspace{4pt}\hspace{7pt}\textbf{Repair example}\vspace{4pt}}}

      \renewcommand{\arraystretch}{1.18}
      \begin{tabularx}{\linewidth}{@{}>{\centering\arraybackslash}p{7mm}@{}X@{}}
        \rowcolor{black!3}[0pt][0pt]
        & \texttt{class PressureCapacity <<requirement>>} \\
        \rowcolor{black!3}[0pt][0pt]
        & \texttt{class ValveModulationRate <<requirement>>} \\
        \rowcolor{black!3}[0pt][0pt]
        & \texttt{derive(ValveModulationRate, PressureCapacity)} \\
        \rowcolor{green!12}[0pt][0pt]
        \textcolor{green!45!black}{\texttt{+}} &
        \texttt{class SupplyVoltageRange <<requirement>>} \\
        \rowcolor{green!12}[0pt][0pt]
        \textcolor{green!45!black}{\texttt{+}} &
        \texttt{satisfy(BrakeControlECU, SupplyVoltageRange)} \\
        \rowcolor{red!10}[0pt][0pt]
        \textcolor{red!70!black}{\texttt{-}} &
        \texttt{derive(ValveModulationRate, PressureCapacity)} \\
        \rowcolor{red!10}[0pt][0pt]
        \textcolor{red!70!black}{\texttt{+}} &
        \texttt{derive(PressureCapacity, ValveModulationRate)} \\
        \rowcolor{red!10}[0pt][0pt]
        \textcolor{red!70!black}{\texttt{!}} &
        \texttt{SupplyVoltageRange remains missing} \\
      \end{tabularx}

      \vspace{2pt}
      \hspace{7pt}\textcolor{green!45!black}{\rule{9pt}{6pt}}\,Correct patch
      \qquad
      \textcolor{red!70!black}{\rule{9pt}{6pt}}\,Model patch
      \vspace{5pt}
    \end{minipage}}

  \vspace{3mm}

  \fcolorbox{black!28}{white}{%
    \begin{minipage}{0.965\textwidth}
      \colorbox{black!8}{%
        \parbox{\dimexpr\linewidth-2\fboxsep\relax}{%
          \vspace{4pt}\hspace{7pt}\textbf{Cross-diagram update example}\vspace{4pt}}}

      \renewcommand{\arraystretch}{1.18}
      \begin{tabularx}{\linewidth}{@{}>{\centering\arraybackslash}p{7mm}@{}X@{}}
        \rowcolor{black!3}[0pt][0pt]
        & \texttt{seed: bdd.puml} \\
        \rowcolor{black!3}[0pt][0pt]
        & \texttt{add(AuxiliarySolenoidInletValves)} \\
        \rowcolor{green!12}[0pt][0pt]
        \textcolor{green!45!black}{\texttt{+}} &
        \texttt{changed\_views: [requirement, sequence]} \\
        \rowcolor{green!12}[0pt][0pt]
        \textcolor{green!45!black}{\texttt{+}} &
        \texttt{requirement: add(AuxiliarySolenoidInletValves)} \\
        \rowcolor{green!12}[0pt][0pt]
        \textcolor{green!45!black}{\texttt{+}} &
        \texttt{sequence: add\_participant(AuxiliarySolenoidInletValves)} \\
        \rowcolor{red!10}[0pt][0pt]
        \textcolor{red!70!black}{\texttt{!}} &
        \texttt{changed\_views: [\,]} \\
        \rowcolor{red!10}[0pt][0pt]
        \textcolor{red!70!black}{\texttt{!}} &
        \texttt{views: \{\}} \\
      \end{tabularx}

      \vspace{2pt}
      \hspace{7pt}\textcolor{green!45!black}{\rule{9pt}{6pt}}\,Correct propagation
      \qquad
      \textcolor{red!70!black}{\rule{9pt}{6pt}}\,Model response
      \vspace{5pt}
    \end{minipage}}

  \caption{Examples of failures in diagram repair and cross-diagram update. Green highlights show
  the expected result, while red highlights show the model output.}
  \label{fig:failure-examples}
\end{figure}

Across the results, failure is rarely caused by invalid syntax. More often, the model either misses
content that should be added, changes the wrong relation, or rewrites valid content while attempting
the repair. These errors can produce professional-looking diagrams, which makes set-level checking
important even when every individual diagram renders.

\subsection{Observations from Human Verification}

Reviewers generally found the systems plausible and coherent across diagrams. The suggested changes
were limited to terminology in the descriptions and did not require changes to diagram code.
Reviewers expressed more uncertainty about notation choices and completeness than about domain
content or cross-diagram consistency. This suggests that automatic checks captured many structural
problems, while human review was most useful for judging modeling conventions, appropriate detail,
and domain language.

Some proposed changes reflected modeling preferences rather than errors. For example, reviewers
interpreted a requirement-relation direction differently, but later confirmed that the
diagram followed the dataset's documented convention. The findings describe the evaluation set and should not
be generalized to the full dataset. More details are provided in Appendix~\ref{app:verification}.

\subsection{Prompt and Model Ablation Studies}

We tested an experimental generation prompt on two fixed contexts using GPT-5.5 and GPT-5.6 Sol.
Both models produced all ten diagrams without a rendering failure or a flagged structural error.
The revised prompt kept requirement relations in their specified direction, produced complete
activity paths, preserved sequence endpoints and message order, and limited state annotations to
information supported by the context. The shared entity anchor made the main structures from both
models very similar. Differences were concentrated in Block Definition properties and operations
and in the amount of detail added to state annotations. GPT-5.5 generally added more properties and
operations, while Sol used fewer annotations and more consistently typed properties in the examined
examples. This small comparison suggests that prompt constraints strongly influence structural consistency,
while model choice affects the amount and form of supporting detail. Details of these experiments are provided in ~\ref{app:development-benchmarks}.

\section{Limitations and Future Work}
\label{sec:limitations}

The human-verification study involved a small number of reviewers and focused on the evaluation 
set. The verified contexts were selected to provide balanced coverage rather than as a
random audit of the complete dataset, so the findings should not be generalized beyond this set.
Future work can expand human review to a larger and more varied sample, refine the rubric, and
involve additional reviewers.

The current dataset represents five diagram types in PlantUML. This provides a practical textual
format, but it does not cover the full range of systems-modeling concepts or tools. Future work
can add more domains, system scales, diagram types, and longer engineering scenarios. They can also
represent models directly in the SysML v2 textual language and validate them with compatible
modeling tools, reducing dependence on PlantUML conventions.

The benchmark evaluates three models from one provider and focuses on diagram repair and
cross-diagram update. This limits conclusions about other model families and engineering tasks.
Future evaluations should include models from other providers, repeated runs, and stronger human
evaluation where more than one output can be correct. The diagram-to-text and diagram-to-diagram
tasks can also be improved and added once their task definitions and evaluation methods reliably
distinguish model performance.

\section{Conclusion}
\label{sec:conclusion}

We introduced \dataset, a dataset of 3,000 engineering contexts and 15,000 connected SysML
diagrams. Each context combines Requirement, Block Definition, Activity, State Machine, and
Sequence diagrams for the same system. The diagram sets were checked for consistency and valid
rendering, and a 100 context for evaluation set was human-verified.
We also introduced benchmarks for diagram repair and cross-diagram update. Evaluation of three
language models shows that syntax errors can usually be repaired, while semantic repair and the
propagation of new information remain difficult. These results demonstrate the need to evaluate
complete diagram sets rather than rendering or isolated diagrams alone. 

\section*{Acknowledgments}

This paper is based on the work supported in part by the industrial sponsors: BEMAC Corporation, ClassNK, and MTICo., Ltd. The authors are grateful for their support and collaboration on this research.
We also thank Sara Sourani, Shih-Chun Deng, Rufus Marcussen, Dezhao Yu, and Charles Wu for helpful discussions and for their assistance during this work.
The authors acknowledge the Center for Advanced Research Computing (CARC) at the University of Southern California for providing computing resources that have contributed to the research results reported within this publication. URL: https://carc.usc.edu.

\bibliographystyle{unsrt}
\bibliography{references}

\clearpage
\appendix
\section{Dataset and Benchmark Details}
\label{app:details}

\subsection{Context contents}

The context is the main unit of \dataset{}. It connects one system description, a shared
model, five diagram sources, rendered previews, validation results, and sources. Table
\ref{tab:appendix-context-fields} summarizes the information stored for each context.

\begin{table}[H]
  \centering
  \caption{Main contents of one dataset context.}
  \label{tab:appendix-context-fields}
  \small
  \begin{tabularx}{\textwidth}{@{}p{0.25\textwidth}Y@{}}
    \toprule
    \textbf{Content} & \textbf{Included information} \\
    \midrule
    Identity & Context identifier, domain, subdomain, and system scale \\
    System description & Components, behavior, operating conditions, requirements, and interfaces \\
    Shared model & Typed entities and relations used to connect the diagrams \\
    Diagram set & Requirement, Block Definition, Activity, State Machine, and Sequence sources \\
    Validation & Schema, rendering, consistency, and file-integrity results \\
    Sources & Source references, model identifiers, prompt versions, and generation details \\
    Verification & Review status, criterion scores, corrections, and resolution details \\
    \bottomrule
  \end{tabularx}
\end{table}

The benchmark split contains 100 contexts, with 25 contexts from each of the four engineering
domains. It contains 36 easy, 30 medium, and 34 hard contexts. Every context contributes the same
five diagram types, giving 500 diagrams in the benchmark set.

\subsection{Dataset statistics}
\label{app:dataset-statistics}

The statistics in this section describe the dataset's 3,000 contexts and 15,000 diagrams, generated
with OpenAI GPT-5.5. Table~\ref{tab:appendix-dataset-coverage}
reports domain and scale coverage.
The dataset spans 244 subdomains. Its scale distribution is close to uniform, while the domain
distribution reflects the available system archetypes and planned variants.

\begin{table}[H]
  \centering
  \caption{Distribution of contexts by engineering domain and system scale.}
  \label{tab:appendix-dataset-coverage}
  \small
  \begin{tabular}{@{}llrr@{}}
    \toprule
    \textbf{Dimension} & \textbf{Category} & \textbf{Contexts} & \textbf{Share} \\
    \midrule
    Domain & Mechanical & 846 & 28.2\% \\
           & Electrical and power & 846 & 28.2\% \\
           & Computer and embedded & 843 & 28.1\% \\
           & Mechatronic and control & 465 & 15.5\% \\
    \midrule
    Scale  & Component & 1,000 & 33.3\% \\
           & Subsystem & 1,002 & 33.4\% \\
           & System & 998 & 33.3\% \\
    \bottomrule
  \end{tabular}
\end{table}

Each context stores one shared entity model before its five diagrams are produced.
Table~\ref{tab:appendix-entity-statistics} reports the number of each entity type. Across the
dataset, these models contain 127,590 entities and 86,578 typed relations. The narrow ranges reflect
the fixed generation requirements, while differences within each range follow the needs of the
described system.

\begin{table}[H]
  \centering
  \caption{Shared entity-model statistics across the 3,000 contexts.}
  \label{tab:appendix-entity-statistics}
  \small
  \begin{tabular}{@{}lrrrr@{}}
    \toprule
    \textbf{Entity type} & \textbf{Total} & \textbf{Mean} & \textbf{Median} & \textbf{Range} \\
    \midrule
    Blocks & 29,290 & 9.76 & 10 & 7 to 11 \\
    Actors & 7,201 & 2.40 & 2 & 1 to 3 \\
    Requirements & 17,810 & 5.94 & 6 & 5 to 7 \\
    States & 21,672 & 7.22 & 7 & 4 to 9 \\
    Activities & 22,707 & 7.57 & 8 & 6 to 8 \\
    Messages & 28,910 & 9.64 & 10 & 8 to 10 \\
    \bottomrule
  \end{tabular}
\end{table}

The five diagram types occur equally because each context contains one of each type.
Table~\ref{tab:appendix-diagram-statistics} gives the size of their source codes. Block
Definition and Requirement diagrams contain more declaration and relation lines, while Activity,
State Machine, and Sequence diagrams use more compact behavioral notation. These values describe
the artifacts and are not used as quality measures.

\begin{table}[H]
  \centering
  \caption{Source codes statistics for each diagram type.}
  \label{tab:appendix-diagram-statistics}
  \small
  \begin{tabular}{@{}lrrr@{}}
    \toprule
    \textbf{Diagram type} & \textbf{Diagrams} & \textbf{Mean lines} & \textbf{Range} \\
    \midrule
    Requirement & 3,000 & 43.74 & 32 to 62 \\
    Block Definition & 3,000 & 65.90 & 43 to 92 \\
    Activity & 3,000 & 16.11 & 14 to 20 \\
    State Machine & 3,000 & 14.01 & 9 to 24 \\
    Sequence & 3,000 & 22.33 & 16 to 30 \\
    \bottomrule
  \end{tabular}
\end{table}

Table~\ref{tab:appendix-generation-statistics} summarizes the generation output and rendering
results. Each diagram type contributes exactly 3,000 diagrams. Nearly every diagram rendered on
the first attempt; and all
failed diagrams rendered successfully after one correction step.

\begin{table}[H]
  \centering
  \caption{Summary statistics for the generated dataset.}
  \label{tab:appendix-generation-statistics}
  \small
  \begin{tabularx}{\textwidth}{@{}p{0.42\textwidth}Y@{}}
    \toprule
    \textbf{Statistic} & \textbf{Value} \\
    \midrule
    Engineering contexts & 3,000 \\
    PlantUML diagrams & 15,000 \\
    Diagrams per context & 5 \\
    Distinct subdomains & 244 \\
    Shared entities & 127,590 \\
    Typed relations & 86,578 \\
    Automatically checked contexts & 3,000 \\
    Human-verified benchmark contexts & 100 \\
    \bottomrule
  \end{tabularx}
\end{table}

\subsection{Example dataset context}

Table~\ref{tab:appendix-example-context} summarizes one human-verified context used in
benchmarks. It describes a four-channel passenger-vehicle ABS hydraulic modulator. The system
description covers normal braking, wheel-lock detection, pressure release and reapplication,
electrical faults, and communication with the driveline controller.

\begin{table}[H]
  \centering
  \caption{Summary of the example ABS context.}
  \label{tab:appendix-example-context}
  \small
  \begin{tabularx}{\textwidth}{@{}p{0.27\textwidth}Y@{}}
    \toprule
    \textbf{Field} & \textbf{Value} \\
    \midrule
    Context identifier & \texttt{ctx-c0004} \\
    System & Four-channel passenger-vehicle ABS hydraulic modulator \\
    Domain and subdomain & Mechanical; automotive brake control \\
    Benchmark difficulty & Medium \\
    Source form & Synthesized from engineering, product, and regulatory references \\
    Shared model & 45 typed entities and 29 typed relations \\
    Diagram set & Five connected diagrams, all rendered successfully \\
    Verification status & \texttt{human-verified} \\
    \bottomrule
  \end{tabularx}
\end{table}

\begin{systemdescription}
The modeled component is a cost-optimized four-channel ABS hydraulic modulator assembly for a
12 V passenger vehicle, installed between a tandem master cylinder and four wheel calipers. The
component contains an aluminum hydraulic modulator block, four normally open solenoid inlet
valves, four normally closed solenoid outlet valves, two low-pressure accumulator chambers, a
pump-back motor driving dual return-pump elements, an embedded brake-control ECU, a brake-pressure
sensor, and a sealed vehicle harness connector. In normal braking, the inlet valves remain open
and the outlet valves remain closed, allowing driver-generated pressure to pass through the block.
During ABS operation, the ECU receives four wheel-speed signals, estimates vehicle reference speed
and slip ratio, and detects wheel lock when wheel behavior exceeds the calibrated 0.10 to 0.30 slip
band. It then closes the affected inlet valve, opens the matching outlet valve, transfers fluid to
the low-pressure accumulator, and starts the pump-back motor before reapplying pressure. The
hydraulic channels shall tolerate 120 bar commanded pressure with a 130 bar short-duration peak,
the electronics shall operate from a 10 to 16 V supply, valve modulation shall support 10 to 15
release and reapply cycles per second, and the packaged assembly shall remain below 1.8 kg. The ECU
also publishes vehicle speed, ABS status, and brake-torque intervention over CAN. After an ECU
reset, the component must return to hydraulic pass-through braking within one control cycle. It
shall identify and indicate wheel-speed sensor, pump-motor, and solenoid-driver faults, inhibit ABS
when closed-loop control is unreliable, and preserve base service braking.
\end{systemdescription}

\begin{figure}[H]
  \centering
  \includegraphics[width=0.64\textwidth]{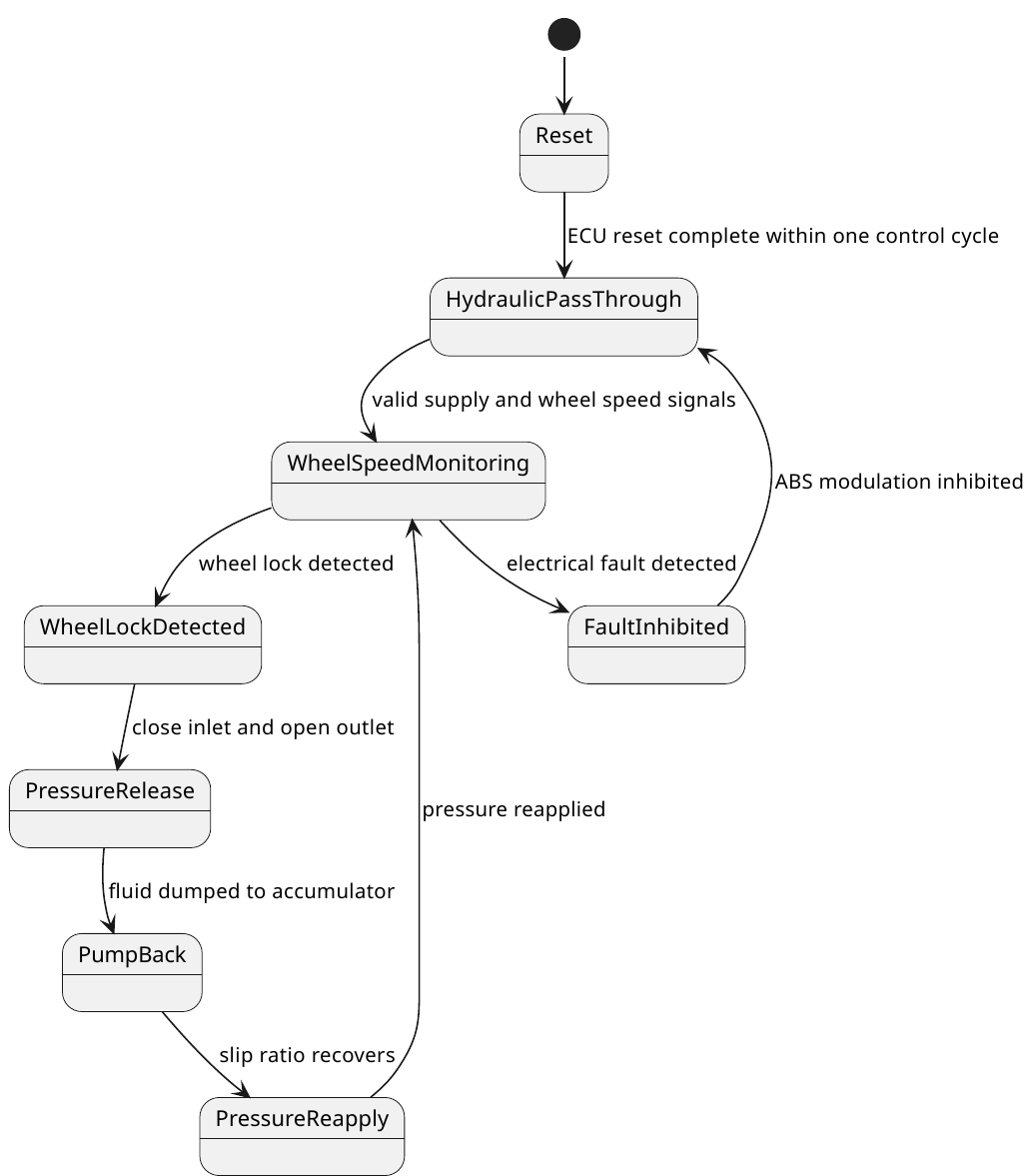}
  \caption{State Machine diagram for the example ABS context.}
  \label{fig:appendix-example-state}
\end{figure}

Figure~\ref{fig:appendix-example-state} shows the State Machine diagram from this context. The
diagram connects normal hydraulic pass-through, wheel-speed monitoring, pressure control, and
fault handling. The same components and behaviors are represented from different perspectives in
the other four diagrams.

\subsection{Benchmark configuration}

Table~\ref{tab:appendix-benchmark-settings} summarizes the shared inference settings and the
task-specific inputs. The reported runs used OpenAI GPT-5.6 Luna, GPT-5.6 Terra, and GPT-5.6 Sol.
All three models received the same examples, prompts, output schema, and settings.

\begin{table}[H]
  \centering
  \caption{Configuration of the reported benchmark runs.}
  \label{tab:appendix-benchmark-settings}
  \small
  \begin{tabularx}{\textwidth}{@{}p{0.28\textwidth}YY@{}}
    \toprule
    \textbf{Setting} & \textbf{Diagram repair} & \textbf{Cross-diagram update} \\
    \midrule
    Provider & OpenAI & OpenAI \\
    Models & GPT-5.6 Luna, GPT-5.6 Terra, and GPT-5.6 Sol & GPT-5.6 Luna, GPT-5.6 Terra,
    and GPT-5.6 Sol \\
    Test contexts & 100 & 100 \\
    Benchmark examples & 1,157 & 300 \\
    Prompting & Zero-shot & Zero-shot \\
    Output & One structured response with complete code & One structured response with selected diagrams and
    complete code \\
    Reasoning effort & Medium & Medium \\
    Responses per example & 1 & 1 \\
    Primary measure & Exact repair & $F_1$ Score \\
    \bottomrule
  \end{tabularx}
\end{table}

\section{Prompts and Models}
\label{app:prompts}

The dataset was created with OpenAI GPT-5.5. The same model was used for the system
descriptions, shared entity models, and five  diagrams. The reported benchmark experiments
used OpenAI GPT-5.6 Luna, GPT-5.6 Terra, and GPT-5.6 Sol. The code blocks below give prompt samples from the ABS context introduced in
Appendix~\ref{app:details}. Long system descriptions and entity lists are shortened where noted.

\subsection{Dataset-generation prompts}

The first prompt grounded each planned context in public engineering information and requested a
structured description. This sample shows the values used for the ABS context.

\begin{promptbox}{Context-generation prompt}
You are researching a real class of engineered system to ground a SysML modeling dataset.

Target system:
Domain: mechanical (brake-control-automotive)
Archetype: ABS braking, four-channel hydraulic, passenger vehicle, cost-optimized
Scale: component
Vocabulary: hydraulic modulator, solenoid inlet valve, solenoid outlet valve, pump-back motor, low-pressure accumulator, driveline integration, wheel speed sensor, slip ratio, wheel lock detection, brake torque

Research how real systems of this class are built. Use at least three independent web sources to identify typical architecture, major components, operating principles, realistic parameters, relevant standards, and common fault or protection behavior.

Write a 300 to 600 word grounding brief with the source URLs. Then write one cohesive 200 to 350 word system description in neutral engineering language. Name the major components, explain how they interact, include three to six realistic quantities with units, and state two to four requirement-like constraints. The description must support structure, activity, state, interaction, and requirement diagrams. Paraphrase the sources and return one structured response containing the description, domain, subdomain, source information, and grounding brief.
\end{promptbox}

The second prompt converted the description into the common model used by all five diagrams.

\begin{promptbox}{Shared entity-model prompt}
You are a systems engineer building a SysML model. From the system description below, extract one canonical set of model elements that every diagram will reference. Return structured data only and do not produce diagram code.

Create 5 to 10 blocks, 1 to 3 actors, 3 to 6 requirements, 4 to 8 states of the main controller, 4 to 8 ordered activities, and 5 to 10 ordered messages. Give every entity a unique short identifier and a fixed diagram-friendly name. Include composition, satisfy, requirement-derivation, and state-transition relations where supported. Every relation endpoint must identify an existing entity. Ground all content in the supplied system description and vocabulary.

System description: {system description}
Domain vocabulary: {seed vocabulary}
\end{promptbox}

The view prompt combined common instructions with the relevant part of the shared entity model and
rules for the requested diagram type. The sample below shows the Requirement view input in compact
form.

\begin{promptbox}{Requirement view-generation prompt}
You generate one SysML Requirement diagram as PlantUML for the system described below. Return one complete PlantUML block and no explanation.

Treat the supplied entity model as read-only. Use only its element names, with the same spelling and case. Do not invent, rename, merge, or omit elements that belong in this diagram. Follow the supplied rules and syntax example for the selected diagram type.

System description:
A four-channel ABS hydraulic modulator controls wheel pressure using inlet and outlet valves, low-pressure accumulators, a pump-back motor, wheel-speed signals, and an embedded brake-control ECU. [Complete description supplied in the prompt.]

Allowed requirements:
PressureCapacity, SupplyVoltageRange, ValveModulationRate, PackagedMassLimit, FailSafePassThrough, ElectricalFaultDiagnostics

Allowed blocks:
ABSHydraulicModulatorAssembly, HydraulicModulatorBlock, SolenoidInletValves, SolenoidOutletValves, BrakeControlECU

Allowed relations:
HydraulicModulatorBlock satisfies PressureCapacity
BrakeControlECU satisfies SupplyVoltageRange
SolenoidInletValves satisfies ValveModulationRate
[Remaining supplied relations omitted here for space.]

Include every requirement and every supplied satisfy or requirement-derivation relation. Include only blocks that satisfy a requirement. Produce the Requirement diagram now.
\end{promptbox}

The diagram-specific instructions were as follows:

\begin{itemize}
  \item \textbf{Requirement:} include every requirement and all applicable satisfy and requirement-
  derivation relations; include only blocks that satisfy a requirement.
  \item \textbf{Block Definition:} include every block and the supplied composition relations; add
  only supported properties and operations.
  \item \textbf{Activity:} include each activity in its given order and one supported decision with
  complete terminal paths.
  \item \textbf{State Machine:} include every supplied state and transition, preserving each trigger
  label and transition direction.
  \item \textbf{Sequence:} include only participating actors and blocks, then add every message in
  its given order with the supplied sender and receiver.
\end{itemize}

If PlantUML rejected a generated diagram, the following repair prompt was added:

\begin{promptbox}{Renderer-repair prompt}
The PlantUML renderer rejected the diagram. Correct only the rendering problem while preserving the diagram type, modeling convention, and supplied entity names. Return one complete corrected PlantUML block and no explanation.

Renderer message: {renderer message}
Rejected source: {PlantUML source}
\end{promptbox}

\subsection{Benchmark prompts}

Every benchmark request used the following system instruction:

\begin{promptbox}{Benchmark system instruction}
You are a careful SysML modelling assistant. Reply with structured output only.
\end{promptbox}

\begin{promptbox}{Diagram-repair prompt}
One view of this SysML diagram set has been damaged. Every other view is correct and unchanged. Restore the damaged view to what it was.

System description: {system description}
Damaged view: {diagram type}
Renderer message, when present: {renderer message}
All five views: {complete PlantUML sources}

Return one structured response containing only the damaged view and its complete PlantUML source. Change only what is wrong. Preserve every unrelated element, relation, and direction. The damage may prevent rendering or may render correctly while contradicting the description and the other four views. Exactly one change caused the damage; find and undo it.
\end{promptbox}

\begin{promptbox}{Cross-diagram update prompt}
You maintain five SysML diagrams that describe the same system. An edit has already been applied to one seed view. Decide which other views must change and return their updated PlantUML.

System description: {system description}
Seed view: {seed diagram type}
Applied edit: {edit description}
Current views: {complete PlantUML sources}

Return one structured response containing a short rationale, the list of changed views, and the complete PlantUML for each listed view. Include a view only when it genuinely references the edited element. Do not return the seed view. Preserve all unrelated content. Remove dangling relations after deletions, and remove an element from a view if its changed type is not valid there. If no other view needs an update, return empty changed-views and views fields.
\end{promptbox}

\section{Additional Results}
\label{app:results}

Main results are reported in the main paper. The following cases give
additional context for the qualitative analysis.

\subsection{Qualitative failure examples}
\label{app:failure-examples}

\paragraph{Repairing an omitted requirement.}
In context \texttt{ctx-c0004}, the semantic repair operator removed the
\texttt{SupplyVoltageRange} requirement and its satisfy relation from the Requirement diagram. The
system description stated that the ABS electronics must operate from a 10 to 16 V supply, and the
Block Definition diagram included the corresponding supply-voltage property. The Sol response did
not restore the requirement. It instead reversed the existing \texttt{deriveReqt} relation between
\texttt{PressureCapacity} and \texttt{ValveModulationRate}. The response therefore rendered, but
failed both defect correction and preservation. This case illustrates why an omission can be hard
to localize: the model must infer absent content from evidence outside the damaged diagram.

\paragraph{Propagating a newly added element.}
In the addition example from the same context, the seed Block Definition diagram introduced
\texttt{AuxiliarySolenoidInletValves} beside the existing inlet-valve block. The expected changes
added the new element to both the Requirement and Sequence diagrams. Luna, Terra, and Sol each
returned an empty list of changed diagrams. Terra and Sol explicitly reasoned that the new block
appeared only in the seed diagram and that existing references remained valid. The responses were
conservative, but they did not infer the requirement coverage and interaction needed for the new
component. The example shows that addition requires reasoning about intended future consistency,
not only matching names already present in the input.

\paragraph{Recovering exchanged participant names.}
Another repair example exchanged \texttt{Driver} and \texttt{SolenoidInletValves} in a Sequence
diagram. This changed the driver into a system participant and made the inlet valves appear as the
source of the brake-pressure input. Luna and Terra returned the damaged diagram without correcting
the exchanged roles. Sol restored both names and their message endpoints. The example is difficult
because every participant and message remains syntactically valid; the error is visible only when
the roles and interactions are compared with the system description and the other diagrams.

\paragraph{Confusing removal with renaming.}
In a cross-diagram update example, the main ABS assembly block was removed from
the seed Block Definition diagram. The expected response removed the corresponding block reference
and its satisfy relation from the Requirement diagram. Luna selected the correct diagram, but
introduced a new V2 assembly block and redirected the relation to the new name.
The response interpreted removal as replacement. It therefore preserved a reference that should
have disappeared and added content not supported by the requested change. This case separates
correct diagram selection from correct update execution.

\section{Development Benchmarks}
\label{app:development-benchmarks}

In addition to the two benchmarks reported in the main evaluation, we tested diagram-to-text
description and diagram-to-diagram translation on development data. Their configurations and
results are summarized here.

\subsection{Diagram-to-text description}

This benchmark gives the model one SysML diagram without the original system description. The
model must write a short description of the system using only the information available in that
diagram. The intended questions are whether a model can understand diagram content and how much
system information is carried by each diagram type.
The benchmark used 25 contexts and created one example for each of the five diagram types,
giving 125 examples per evaluated model. An independent language model rated a balanced subset of
200 responses on system identity, structural content, and writing form using a five-point scale.
Reference descriptions and unrelated descriptions were included to check the upper and lower
parts of the scale. The tested models scored between 4.84 and 4.93 on the five-point scale. The per-view results also showed that Activity and State Machine diagrams carry less information about the complete system than Requirement and Block Definition diagrams.

\subsection{Diagram-to-diagram translation}

This benchmark gives the model one diagram and asks it to create a different diagram for the same
system. This test examined six directions among Requirement, Block Definition, and Sequence
diagrams. It used 25 contexts, producing 150 translation examples for each evaluated
model. Four models were tested across two stages.
Because a source diagram can support several target diagrams, the test used an independent
language model to rate the returned elements, relations, and output form. Reference outputs,
copied source diagrams, and unrelated diagrams were included to check the rating scale.
The model averages ranged from 3.72 to 3.88 on the five-point scale. Scores varied by translation
direction because each source diagram exposes a different part of the system. Requirement, Block
Definition, and Sequence diagrams therefore contributed different evidence to the generated target.

\section{Human Verification Materials}
\label{app:verification}

Two reviewer roles each completed 60 set-level reviews. Twenty contexts were reviewed by both roles,
giving 120 completed reviews for 100 unique contexts. Reviewers checked rendering, SysML correctness,
domain plausibility, cross-diagram consistency, and completeness. Every accepted criterion score
was at least 2 on the 0 to 3 rubric.
The initial reviews contained 53 recommendations to verify without changes and 67 recommendations
for minor correction, affecting 64 unique contexts. Adjudication applied only two terminology
corrections. One replaced an aircraft-specific qualification term in a spacecraft context. The
other replaced an inapplicable automotive safety reference with machinery-safety framing. Neither
change modified the diagrams, and no context was rejected or replaced.
Table~\ref{tab:appendix-verification-summary} reports the final scores. All 500 diagrams
were rendered again after resolution and all rendered successfully.

\begin{table}[H]
  \centering
  \caption{Final human-verification scores for the benchmark core.}
  \label{tab:appendix-verification-summary}
  \small
  \begin{tabular}{@{}lrr@{}}
    \toprule
    \textbf{Criterion} & \textbf{Score 2} & \textbf{Score 3} \\
    \midrule
    SysML correctness & 60 & 40 \\
    Domain plausibility & 2 & 98 \\
    Cross-diagram consistency & 2 & 98 \\
    Completeness & 29 & 71 \\
    \bottomrule
  \end{tabular}
\end{table}

\end{document}